\documentstyle[epsfig,prl,twocolumn,aps]{revtex}
\input{psfig}

\begin{document}
\twocolumn[\hsize\textwidth\columnwidth\hsize\csname @twocolumnfalse\endcsname
\title{Continuous Emission versus Freeze-out via HBT}
\author{Fr\'ed\'erique Grassi$^1$, Yogiro Hama$^1$,  
        Sandra S.\ Padula$^2$ and Otavio Socolowski Jr.$^1$} 
\address{$^1$ Instituto de F\'{\i}sica, Universidade de S\~ao Paulo, 
         CP 66318,\\
         05315-970 S\~ao Paulo, Brazil.}
\address{$^2$ Instituto de F\'{\i}sica Te\'orica, 
         Universidade  Estadual Paulista, \\  
         Rua Pamplona 145, CEP 01405-900 S\~ao Paulo, Brazil.}
\maketitle
\begin{abstract}

The effect of the {\bf continuous emission} hypothesis on the two-pion 
Bose-Einstein correlation function is discussed and compared with the 
corresponding results based on the usual {\bf freeze-out}. Sizeable 
differences in the correlation function appear in these different 
descriptions of the decoupling process. This means that, when extracting 
properties of the hot matter formed in high-energy heavy-ion collisions 
from the data, completely different conclusions may be reached according 
to the description of the particle emission process adopted. 

\vspace{2pc}


\end{abstract}
\vskip2pc]


\section{Introduction}

When describing ultra-relativistic heavy-ion collisions with hydrodynamic 
models, a simple picture has been extensively adopted. It is usually 
considered that, as the thermalized matter expands, the system gradually 
cools down and, when the temperature reaches a certain freeze-out value 
$T_f$, it decouples. Every observed quantity is then computed on the 
hypersurface $T=T_f$. For instance, the momentum distribution of the 
produced hadrons are obtained by using the Cooper-Frye integral\cite{CF} 
extended over this hypersurface. Though operationally simple, such a 
zero-thickness freeze-out hypersurface is clearly a highly idealized 
concept when applied to finite-volume and finite-lifetime systems as 
those formed in high-energy heavy-ion collisions.

More recently, Grassi, Hama and Kodama\cite{ghk} proposed an alternative
picture to the particle emission: instead of being emitted only when
crossing the sharply defined freeze-out surface, they considered that the
process could occur continuously. Being so, in this picture, particles 
could be emitted from the whole expanding volume of the system, at 
different temperatures, and not only from the surface with constant 
$T=T_f$. As a consequence, in the continuous emission model (CEM), the 
observed quantities depend on the whole history of the expanding system 
and not only on the instant of the freeze-out. Concretely, it has been 
shown\cite{ghk,ghks} that i) CEM enhances the large$-m_T$ component of 
the heavy-particle ($p, \Lambda, \Xi, \Omega ,...$) $m_T$ spectra, ii) it 
gives a concave shape for the pion $m_T$ spectrum even without 
considering transverse expansion of the fluid, iii) it can lead to the 
correct hyperon production ratios and spectrum shapes with 
conceptually reasonable choice of parameters\cite{ghk,ghks,gs1}, and 
iv) it reproduces the observed mass dependence of the slope parameter $T$ 
\cite{gs2}. 

Naturally, we would like to further explore if the above model would 
present striking differences when compared to the usual sudden freeze-out 
picture. One expectation would be that the space-time region from which 
the particles were emitted would be quite different in both scenarios. In 
the continuous emission picture the duration of the emission processes is 
expected to be longer than in the freeze-out scenario, which should 
considerably affect the behavior of the correlation function. Previous 
studies have indeed shown that the influence of the emission time 
\cite{hamapad,pratthpb,gyupa89,pg:nioc} on the apparent transverse source 
dimensions were remarkably strong. It was also shown in 
Ref.\cite{gyupa89,pg:nioc} that a prolonged freeze-out would considerably 
distort the two-particle correlation function. Our main object in the 
present work is to show the differences in two-pion correlation predicted 
by CEM, as compared with the results obtained under the usual assumption 
of sharp freeze-out. For this purpose, we will adopt the same 
approximations used in Ref. \cite{ghk}, namely, one dimensional Bjorken 
model\cite{Bj} for massless-pion gas. It turns out that, within these 
approximations, the HBT effect suffers a large deformation when the usual 
freezeout scenario is replaced by CEM, affecting substantially the 
conclusions achieved on the properties of the matter formed in 
high-energy collisions. 

\section{Continuous emission of particles}

In CEM, it is assumed that, at each space-time point $x^\mu $, each 
particle has a certain probability of not colliding any more, due to the 
finite dimensions and lifetime of the thermalized matter. Then, the 
distribution function $f(x,p)$ of the expanding system has two components, 
one representing the portion of the fluid already free and another 
corresponding to the part still interacting, i.e., 
\[
f(x,p)=f_{free}(x,p)+f_{int}(x,p)\,. 
\]
We may write the portion of free particles as a fraction of the total
distribution function, as follows 
\begin{equation}
f_{free}(x,p)={\cal P}f(x,p)=\frac{{\cal P}}{1-{\cal P}}f_{int}(x,p)\;.
\label{ffree}
\end{equation}
Let us assume, as in the previous papers, that the fraction still
interacting is represented by a thermal distribution function 
\begin{equation}
f_{int}(x,p)\approx f_{th}(x,p)=\frac g{(2\pi )^3}\frac 1{\exp {\left[
p.u(x)/T(x)\right] }\pm 1}\;,  \label{fth}
\end{equation}
where $u^\mu $ is the fluid velocity at $x^\mu $ and $T$ is its 
temperature at that point. The factor ${\cal P}$ can be alternatively 
understood as the probability that a particle with momentum $p^\mu $ 
escapes from $x^\mu $ without further collisions.

If we assume that the fluid is confined to a cylinder of radius 
$R_T$\thinspace , the fraction ${\cal P}$ of free particles at each 
space-time point $x^\mu $ may be computed by using the Glauber formula 
\begin{equation}
{\cal P}=\exp {\ \left( -\int_t^{t_{out}}n(x^{\prime })\sigma
v_{rel}dt^{\prime }\right) }\;\;,  \label{P}
\end{equation}
where

\begin{equation}
t_{out}=t+(-\rho \cos \phi +\sqrt{R_T^2-\rho ^2\sin ^2\phi })/(v\sin \theta) 
\label{tout}
\end{equation}
is the time when the particle with velocity 
$\vec{v}=(v\sin \theta \cos \phi \,,\,v\sin \theta \sin \phi \,,v\cos \theta )$ 
reaches the surface of the fluid at $\rho=R_T\,$.

If we further consider that, initially, the energy density is 
approximately constant (i.e., $\epsilon =\frac{\pi ^2}{10}T_o^4$ for all 
the points with $\rho \le R_T$ and zero for $\rho >R_T$), we can 
calculate the probability ${\cal P}$ analytically, resulting in 

\begin{equation}
{\cal P}=(\tau /\tau _{out})^a;\;\;a\sim 3\frac{1.202}{\pi ^2}T_0^3\tau
\sigma v_{rel}\;\;,  \label{PGl}
\end{equation}
where $v_{rel}\approx 1$. The previous results can be found in Ref.\cite
{ghk,ghks,gs1,gs2}.

\section{HBT interferometry}

The second-order interferometry of identical particles, also known as HBT 
effect\cite{zajc:boal} is a powerful tool for probing geometrical sizes 
of the space-time zone from which they were emitted, as well as for 
testing dynamical correlations built in during the system evolution.

In its idealized version, the two-pion interferometry could be studied 
through the so-called two-particle correlation function 
\begin{eqnarray}
C_2(k_1,k_2)=\frac{P_2(k_1,k_2)}{P_1(k_1)P_1(k_2)}=1+|\rho (k_1-k_2)|^2\;\;,
\label{w}
\end{eqnarray}
where $P_1(k_i)$ and $P_2(k_1,k_2)$ are, respectively, the 
single-particle inclusive distribution and the joint probability for 
detecting two pions; $\rho (k_1-k_2)$ is the Fourier transform of the 
source space-time distribution.

In realistic cases, however, it is mandatory to employ more general 
formalisms\cite{hamapad,pratthpb,gyupa89,pg:nioc,ccef}, as is the case of 
the Covariant Current Ensemble, flexible enough to include phase-space 
correlations resulting from the underlying dynamics. As a consequence, 
the HBT correlation functions would reflect a model dependent analysis. 
In the Covariant Current Ensemble formalism, the correlation function can 
be expressed as\cite{pg:nioc,ccef} 
\begin{equation}
C(k_1,k_2)=C(q,K)=1+\frac{|G(q,K)|^2}{G(k_1,k_1)G(k_2,k_2)}\;\;,  \label{cce}
\end{equation}
where $q^\mu =k_1^\mu -k_2^\mu $ and $K^\mu =\frac 12(k_1^\mu +k_2^\mu )$
and the complex amplitude, $G(k_1,k_2)$, can be written as 
\begin{equation}
G(k_1,k_2)=\int d^4xd^4p\;e^{iq^\mu x_\mu }D(x,p)j_0^{*}(u_f^\mu k_{1\mu
})j_0(u_f^\mu k_{2\mu })\;\;,  \label{g12}
\end{equation}
where $D(x,p)$ is the break-up phase-space 
distribution\cite{gyupa89,pg:nioc,ccef} and the currents, $j_0(u_f.k_i)$, 
contain information about the production dynamics. If one takes $k_1=k_2$ 
in Eq. (\ref{g12}), one obtains 
\begin{equation}
G(k_i,k_i)=\int d^4xd^4p\;D(x,p)|j_0(u_f^\mu k_{i\mu })|^2\;\;, \label{gii}
\end{equation}
which coincides with the one-particle spectrum.

As discussed in Ref.\cite{ccef}, the currents $j_0(u_f.k_i)$ in Eqs. 
(\ref{g12},\ref{gii}) can be associated to thermal models and written 
covariantly as $j_0(k)\propto\sqrt{u^\mu k_\mu}\exp{\{-u^\mu k_\mu/(2T)\}}$. 
However, to make the computation easier, we shall adopt throughout the 
paper a more convenient parametrization 
\begin{equation}
j_0(u.k)=\exp {\{-\frac{u^\mu k_\mu }{2T_{ps}}\}}\;\;,  \label{j0pseudo}
\end{equation}
where, in the case of pions, the so-called pseudo temperature $T_{ps}$ 
was related with the true temperature $T$ by the equation\cite{ccef} 
\begin{equation}
T_{ps}(x)=1.42\,T(x)-12.7\text{ MeV\ }.  \label{tps}
\end{equation}
This mapping between $T(x)$ and $T_{ps}(x)$ was later shown to be a good
approximation also in the case of kaon interferometry\cite{padrol}.

\subsection{Bjorken model with sudden freezeout}

In the ideal one dimensional Bjorken picture, using the above 
pseudo-thermal parameterization for the currents, an analytical form for 
the amplitudes can be derived \cite{ccef} 
\begin{equation}
G(k_1,k_2)=2<\frac{dN}{dy}>\{\frac 2{q_TR_T}J_1(q_TR_T)\}K_0(\xi )\;\;,
\label{gk1k2}
\end{equation}
where 
\begin{eqnarray}
\xi^2&=&[\frac 1{2T}(m_{1T}+m_{2T})-i\tau(m_{1T}-m_{2T})]^2+ \nonumber\\
&&2\,(\frac 1{4T^2}+\tau ^2)\,m_{1T}\,m_{2T}\,[\cosh (\Delta y)-1]\;\;,
\label{xi}
\end{eqnarray}
$\Delta y=y_1-y_2$ and $<\ >$ indicates average over particles 1 and 2. 

The single-inclusive distribution is then written as 
\begin{equation}
G(k_i,k_i)=E\frac{d^3N}{dk_i^3}=2\frac{dN}{dy_i}K_0(\frac{m_{iT}}T)\;.
\label{gkk}
\end{equation}

\subsection{Bjorken model with continuous emission}

The initial expectation concerning the differences between the continuous
emission versus the freeze-out scenarios were mainly focused on the
different emission periods. Naturally, in the continuous emission picture
the duration of the emission processes is longer than in the freeze-out
scenario, which should considerably affect the behavior of the correlation
function.The reason for this comes from previous studies which have shown 
that the influence of the emission time \cite{hamapad,pratthpb,gyupa89,pg:nioc} on the transverse source 
dimensions were remarkably strong.

For treating pion interferometry in the case that interests us, we 
consider a different but equivalent form for expressing the amplitudes in 
Eq. (\ref{cce}). The single-inclusive distribution is written as in 
Ref.\cite{ghk} 
\begin{equation}
G(k_i,k_i)=\int d^4x\;{\cal D}_\mu \left[ k_i^\mu f_{free}\right] \;\;.
\label{giice}
\end{equation}

Analogously, the two-particle complex amplitude is written, instead of 
Eq. (\ref{g12}), as 
\begin{equation}
G(k_1,k_2) = \int d^{4}x e^{i q x} \; \{ {\cal D}_\mu \left[ k_1^\mu
f_{free} \right] \}^\frac{1}{2} \; \{ {\cal D}_\mu \left[ k_2^\mu f_{free}
\right] \}^\frac{1}{2} \; \; .  \label{g12ce}
\end{equation}

In Eq.(\ref{giice}) and (\ref{g12ce}), ${\cal D}_\mu $ is the generalized
divergence operator, which, due to the symmetry of the problem, is 
written in Bjorken+transverse polar coordinates.

In order to proceed further, let us recall that usually we are interested 
in small momentum differences $q^\mu =k_1^\mu -k_2^\mu \,$, as compared 
with the {\it average momentum of the pair}, 
$K^\mu =\frac 12(k_1^\mu +k_2^\mu )$. 
If we then approximate $k_i^\mu \approx K^\mu $ in Eq.(\ref{g12ce}), a
substantial simplification is achieved and it could then be written as 
\begin{equation}
G(q,K)\equiv G(k_1,k_2)=\int d^4x\,e^{iq^\nu x_\nu }\,{\cal D}_\mu\left[
K^\mu f_{free}\right] \;.  \label{g12Kce}
\end{equation}
We should note that such a dependence on $K^\mu$, replacing the 
individual momenta $k_1^\mu$ and $k_2^\mu$ in the complex amplitude of   
Eq.(\ref{g12ce}), is also present in the general derivations based on 
the Wigner formalism. 

In principle, the integral in Eq. (\ref{g12Kce}) should be extended over 
the whole space-time with $\tau >\tau _0$\thinspace. However, due to the 
finite size and lifetime of our system, the integrand is expected to 
quickly vanish where the assumption embodied by Eq. (\ref{fth}) also 
breaks down. So, in computing this integral, we separated the space-time 
in two regions, one where ${\bf {\cal P}}>{\bf {\cal P}}_{{\cal F}}$ and 
the other with ${\bf {\cal P}}\leq {\bf {\cal P}}_{{\cal F}}\,$, with 
some reasonable value of ${\bf {\cal P}}_{{\cal F}}\,$. Upon partial 
integration, the latter is reduced to the surface contribution and the 
former may be estimated by using the Cooper-Frye formula\cite{CF} on the 
surface ${\bf {\cal P}}={\bf {\cal P}}_{{\cal F}}\,$, applied to the 
interacting component. We emphasize, however, that ${\bf {\cal P}}$ is a 
momentum-dependent quantity, so this is not the usual Cooper-Frye 
integral. After some manipulation, we get for the single-inclusive 
distribution 

\begin{eqnarray}
G(k_i,k_i) &=&\frac 1{(2\pi )^3(1-{\bf {\cal P}}_{{\cal F}})}\int_0^{2\pi
}d\phi \int_{-\infty }^{+\infty }d\eta  \nonumber \\
&\times &\{\int_0^{R_T}\rho \,d\rho \;\tau _{{\cal F}}\;m_{iT}\,
\cosh(y_i-\eta )  \nonumber \\
&+&\int_{\tau _0}^{+\infty}\!\!\tau \,d\tau \rho _{{\cal F}}k_{iT}\cos \phi
\}e^{-m_{iT}\cosh (y_i-\eta )/T_{ps}(x)}.  \nonumber \\
&&  \label{giicontemis}
\end{eqnarray}

Analogously, instead of Eq. (\ref{gk1k2}), the two-particle complex
amplitude is now written as 
\begin{eqnarray}
G(q,K) &=&\frac 1{(2\pi )^3(1-{\bf {\cal P}}_{{\cal F}})}\int_0^{2\pi }d\phi
\int_{-\infty }^{+\infty }d\eta  \nonumber \\
&\times &\{\int_0^{R_T}\rho \;d\rho \;\tau _{{\cal F}}\;M_T\;\cosh (Y-\eta) 
\nonumber \\
&&\times e^{i[\tau _{{\cal F}}(q_0\cosh \eta -q_L\sinh \eta )-\rho q_T\cos
(\phi -\phi _q)]}  \nonumber \\
&+&\!\int_{\tau _0}^{+\infty }\!\!\tau d\tau \rho _{{\cal F}}K_T\cos \phi 
\nonumber \\
&&\times \,e^{i[\tau (q_0\cosh \eta -q_L\sinh \eta )-\rho _{{\cal F}}q_T\cos
(\phi -\phi _q)]}\}  \nonumber \\
&\times & e^{-M_T\cosh (Y-\eta )/T_{ps}(x)},  \label{g12contemis}
\end{eqnarray}
where \\{\small \smallskip{$M_T=\!\sqrt{K_T^2+M^2},\;\vec{K}_T=\frac 12 
(\vec{k}_1+\vec{k}_2)_T\,,\,M^2=K_\mu K^\mu=m^2-\frac 14q_\mu q^\mu $},} 
$Y$ is the rapidity corresponding to $\vec{K}$, $\phi $ is the azimuthal 
angle with respect to the direction of $\vec{K}$, and $\phi _q$ is the 
angle between the directions of $\vec{q}$ and $\vec{K}$.

In Eqs. (\ref{giicontemis}) and (\ref{g12contemis}), $\tau _{{\cal F}}$ 
and $\rho _{{\cal F}}$ are the limiting values corresponding to a 
certain value of the escape probability ${\cal P}_{{\cal F}}\,$, 
{\it i.e.},

{\small 
\begin{eqnarray}
\tau _{{\cal F}} &=&\frac{-\rho\cos\phi+\sqrt{R_T^2-\rho ^2\sin ^2\phi }}
{(k_T/E)\cosh y\,\bigl[\sqrt{\sinh ^2(\eta -y)+
{\bf {\cal P}}_{{\cal F}}^{-2/a}}-\cosh (\eta -y)\bigr]}\;.  \nonumber  
\label{tauf} \\
&&  \label{tauf}
\end{eqnarray}
} and \\ {\small 
\begin{eqnarray}
\rho_{{\cal F}} &=&-\tau \frac{k_T}E\cosh y\cos \phi \bigl[\sqrt{\sinh^2 
(\eta-y)+{\bf {\cal P}}_{{\cal F}}^{-2/a}}\!\!-\cosh(\eta-y)\bigr] 
\nonumber \\
&\pm &\Bigl[ R_T^2-\tau ^2(\frac{k_T}E)^2\cosh ^2y\sin ^2\!\phi \,
\bigl[\sqrt{\sinh ^2\!(\eta -y)+{\bf {\cal P}}_{{\cal F}}^{-2/a}} \nonumber \\
&&-\cosh (\eta -y)\bigr]^2\Bigr]^{1/2}.  \label{rhof}
\end{eqnarray}
}

For choosing the value of ${\cal P}_{{\cal F}}\,$, in principle, we would
like to take ${\cal P}_{{\cal F}}=1$, corresponding to the complete
integration of (\ref{giice}) and (\ref{g12Kce}). However, we should notice 
that the expressions (\ref{giicontemis}) and (\ref{g12contemis}) above 
become indeterminate in the limit ${\cal P}_{{\cal F}}\rightarrow 1$. As 
mentioned above, the thermal assumption for $f_{int}(x,p)$ breaks down in 
the same limit. For this reason, already in Ref.\cite{ghk}, it was chosen 
${\bf {\cal P}}_{{\cal F}}=0.5$ and the effect of changing this value was 
discussed. We shall adopt the same value here. 

\bigskip

\section{Comparison of Results}

\subsection{Ideal Configurations}

The complexity of the expressions for the amplitudes appearing in Eq. 
(\ref{cce}) in the continuous-emission scenario is evident from 
Eq.(\ref{giicontemis}) and (\ref{g12contemis}) above. In order to get 
some insight regarding the differences of the correlation functions in 
the two scenarios under investigation, let us select some special 
kinematical zones, corresponding to an idealized situation in which high 
precision data with {\sl unlimited} statistics would be available. For 
instance, let us fix $y_i=0$ ($K_L=q_L=0$), so that $\theta =\pi /2$ with 
respect to the collision axis. Due to the symmetry of the problem we can, 
without any loss of generality, choose $\vec{K}$ along the $x$-axis. We 
then explore the behavior of $C(q_T,K_T)$ for fixed {\bf $K_T$}. For 
restricting even more our kinematical window, let us consider two cases. 
Case I (or Zone I) corresponds to considering 
$\phi_{p_1}=-\phi_{p_2}=\phi_p$ (the two pions are symmetrically emitted 
around $\vec{K}$), implying that $\phi _q=\pi /2$; in this case, we can 
write the individual momenta as 
$k_i^\mu=(\sqrt{m_\pi^2+K_T^2+q_T^2/4}\;,\;K_T\;,\;\pm q_T/2\;,\;0\,)$, 
where the $\pm $ signs correspond to pion 1 and 2, respectively. The 
momentum difference $\vec{q}_T=\vec{q}_S$ in this situation corresponds 
to the so-called {\sl sidewards} component introduced in 
Ref\cite{hamapad,pratthpb}. For comparison, we consider that in the usual 
freezeout scenario, the decoupling occurs at $T_{fo}=170$ MeV. The other 
constant values assumed in the calculation that follows were: \bigskip

\begin{tabular}{|c|c|c|c|c|}
\hline
$T_0$ & $\tau_0$ & $<\sigma v_{rel}>$ & $R_T$             & $m_\pi$ \\ 
(MeV) & (fm/c)   & (fm$^2$)           & (fm)              & (MeV)\\ \hline
200   & 1        & 2                  & 3.7 ($\approx$ S) & 140 \\ \hline
\end{tabular}
\bigskip

\begin{figure}[h]\epsfxsize=7.2cm
\centerline{\epsfbox{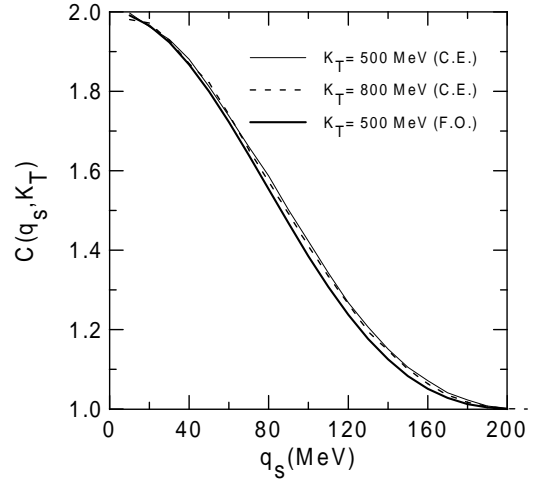}} 
\caption{\it \small $\pi \pi $ Correlation Function {\sl x} $q_T$ in 
Kinematical Zone I ({\sl side} direction), showing the curves 
corresponding to the continuous emission with that corresponding to the 
sudden freezeout. This last case is not sensitive to $K_T$ in this 
kinematical zone but a slight dependence on $K_T$ can be seen for 
continuous emission.} 
\label{fig1.eps}
\end{figure}

Results corresponding to the ZONE I above are shown in Fig. 1. As 
expected, since we are neglecting the transverse expansion, the 
difference between the predictions of the two scenarios is small. 
Slightly broader correlation function for the CEM case, and the 
decreasing width with $K_T\,$, was also expected.

Case II (or Zone II) correponds to considering $\phi_q=0$, 
$|\vec{k}_1|>|\vec{k}_2|\,$, with both $\vec{k}_i$ along the 
$x$-axis, {\it i.e.}, 
$\vec{k}_i\parallel\vec{K}_T \parallel \vec{q}_T$; in this case, we can 
write the individual momenta as {\small $k_i^\mu =(\sqrt{m_\pi^2+
[K_T\pm q_T/2]^2},K_T\pm q_T/2\,,0\,,0)$}, where again the $\pm$
signs correspond to pion 1 and 2, respectively. The momentum difference 
$\vec{q}_T=\vec{q}_O$ in this situation corresponds to the so-called 
{\sl outwards} compo-\hfilneg\  

\begin{figure}[h]\epsfxsize=7.2cm
\centerline{\epsfbox{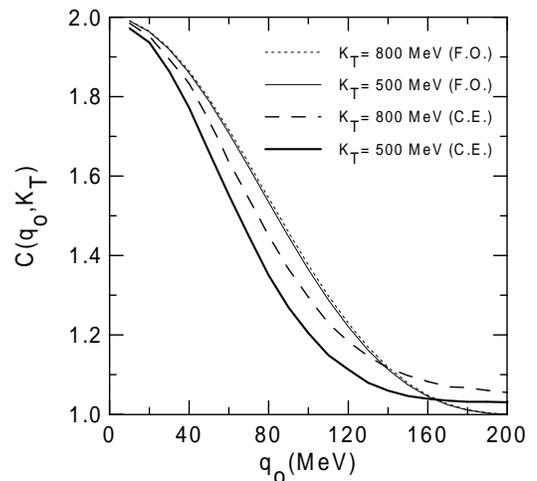}} 
\caption{\it \small $\pi \pi $ Correlation Function {\sl x} $q_T$ in 
Kinematical Zone II ({\sl out} direction), showing the curves 
corresponding to the continuous emission with the ones corresponding to 
the sudden freezeout.} 
\label{fig2.eps}
\end{figure}

\noindent 
nent introduced in Ref\cite{hamapad,pratthpb}. Results
corresponding to the ZONE II above are shown in Fig. 2. This is the case,
mentioned in the Introduction, where the duration of the emission process
becomes essential\cite{hamapad,pratthpb,gyupa89,pg:nioc}. Since the 
emission time in the usual freezeout does not depend crucially on the 
particle momentum, the correlation function is almost independent of 
$K_T\,$. On the contrary, the emission time is strongly momentum 
dependent in\hfilneg\ 

\noindent
CEM, for large-momentum particles are emitted mainly at early times, 
whereas small-momentum particles may be emitted also at later stage of 
expansion when the fluid is cooler and the system larger. So, we see in 
Fig. 2 a significant $K_T\,$ dependence, being the correlation narrower 
for smaller $K_T\,$. Also, we can see that both the CEM curves are 
narrower than the corresponding freezeout curves, indicating that the 
emission time in CEM is longer in general. 
Looking more carefully at the curves, we can also perceive that the tail 
of $C(q_T,K_T)$ in CEM is much flatter than in sharp freezeout. Probably 
this flat tail is due to the small source depth in CEM at early times. It 
is clear that in Bjorken model without transverse expansion, which we 
used in the present work, the source depth is constant and $\sim R_T$ in 
sharp freezeout scenario. 

Figure 3 represents the same situation as in Zone II but with a different 
freezeout temperature, $T_{fo}=140$ MeV, in order to show the sensitivity 
of the results for a lower $T_{fo}$ in the case of the usual freezeout. 
As seen, the $T_{fo}\,$ dependence has shown to be very weak, as 
expected.

\begin{figure}[h]\epsfxsize=7.2cm 
\centerline{\epsfbox{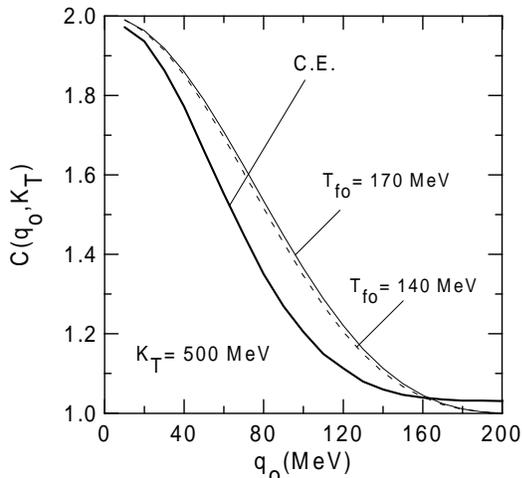}} 
\caption{\it \small Study of the {sensitivity to $T_{fo}$ } in the abrupt 
freezeout model in the Kinematical Zone II ({\sl out} direction). 
Together with previous results shown in Fig.2, the curve corresponding to 
a lower freezeout temperature, $T_{fo}=140 MeV$, is also included.} 
\label{fig3.eps}
\end{figure}

\subsection{Averaged Correlations}

Although the selective kinematical zones could teach us interesting 
points concerning the differences of the behavior of the correlation 
functions corresponding to both scenarios, as shown in Fig. 1-3, such 
conditions correspond to an idealization. For putting the calculations 
into more realistic grounds, averages over the angles, momenta, and the 
unobserved projections of the momentum differences $\vec{q}$ should be 
performed. Using the azimuthal symmetry of the problem we can still 
select $\vec{K}_T$ along the $x$-axis, such that $\vec{K}=(K_T,0,K_L)$. 
Then, averaging over different kinematical zones or windows would 
correspond to integrating over $\vec{K}$ and $\vec{q}$ (except over the 
plotting component of $\vec{q}$). In order to make the analysis roughly 
compatible with the range covered by NA35 S+A collisions\cite{alber}, we 
considered the kinematical variables in the following intervals: 
$-0.5\leq y\leq 0.5$ (or, equivalently, $-180\leq K_L\leq 180$ MeV); 
$50\leq K_T\leq 600$ MeV; $0\leq (q_L,q_S,q_{out})\leq 30$ MeV 
(corresponding to the first experimental bin). As an illustration, we 
show below an example about how to compute the average:

{\small 
\begin{eqnarray}
&&\langle C(q_L)\rangle =1+  \nonumber \\
&&\frac{\int_{-180}^{180}dK_L\int_{50}^{600}dK_T\int_0^{30}dq_S
  \int_0^{30}dq_oC(K,q)|G(K,q)|^2}{\int_{-180}^{180}dK_L 
  \!\int_{50}^{600}dK_T\!\int_0^{30}dq_S\!\int_0^{30}dq_o   
  C(K,q)G(k_1,k_1)G(k_2,k_2)}\,.  \nonumber
\end{eqnarray}
}

The results are presented in the following way. First, as done in the 
preceding subsection, in order to stress the differences of results 
predicted by the two scenarios under study, we start from the same 
initial temperature $T_0=200$ MeV for both the usual freezeout and CEM. 
The results are shown in Figs. 4-6, respectively as functions of 
$q_L\,$, $q_O$ and $q_S$. One sees in Fig. 4 that, as is well known, the 
$q_L$ dependence is very sensitive to the freeze-out temperature 
$T_{fo}$ and if the {\bf same} initial temperature is attained in both 
scenarios, the correlation function cor-\hfilneg\ 

~ \vspace{-.3cm}

\begin{figure}[h]\epsfxsize=8cm
\centerline{\epsfbox{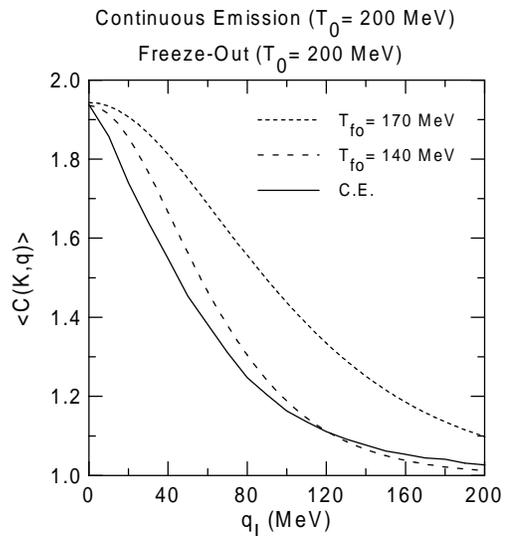}} 
\caption{\it \small $\langle$ C $\rangle_{\pi \pi}$ {\sl x} $q_L$ 
averaged over $q_S$ and $q_{o}$, showing the curve corresponding to the 
continuous emission hypothesis with the ones corresponding to the usual 
freezeout for $T_{fo}=170$ MeV and $T_{fo}=140$ MeV.} 
\label{fig4.eps}
\end{figure}

\noindent
responding to the continuous 
emission picture is closer to the one referring to the thermal freezeout 
at lower $T_{fo}$. However, the shapes are not the same. The one related 
to CEM is more peaked at the small-$q_L$ values, becoming flatter in the 
tail region. This is in clear contrast to those corresponding to sharp 
freezeout scenario which are more similar to Gaussians. Physically, this 
behavior of CEM curve could be interpreted as exhibiting the history 
of 
%
the matter in expansion, because particles are emitted during the whole 
evolution in CEM. Namely, the tail of $\langle$C$\rangle$ depends 
essentially on early times, when the size of the fluid is small and 
its temperature high, whereas the peak reflects later times, 
when the fluid has fully expanded and cooled down. 

We have already seen in the preceding subsection that, when plotted as 
function of $q_O\,$, the correlation function in CEM is significantly
narrower than the one in the sharp freezeout and has a flatter tail. 
These features are again seen in Fig. 5, where 
$\langle C\rangle _{\pi \pi }$ with lower freeze-out temperature 
$T_{fo}\,$ is closer to the curve for CEM, as in Fig. 4. However, if the 
{\bf same} initial temperature is attainded in both scenarios, very low 
$T_{fo}\,$ is necessary, in this case, in order to approximately 
reproduce the same correlation predicted by CEM. We can also notice that 
the depletion of the correlation function at small $q_O$ values is more 
dramatic in CEM. In any case, it is important to emphasize that 
our source is totally chaotic in both scenarios and, as is well 
known\cite{hamapad}, $\langle C\rangle <2$ at $q_L=0$ is originated only 
from the averaging processes, since Coulomb final state interactions, as 
well as the effect of resonances decaying into 
$\pi$'s\cite{gyupa89,pg:nioc} were not considered here.

We can again notice in Fig. 6 that, also as function of\hfilneg\ 

~ \vspace{-.3cm}

\begin{figure}[h]\epsfxsize=8cm
\centerline{\epsfbox{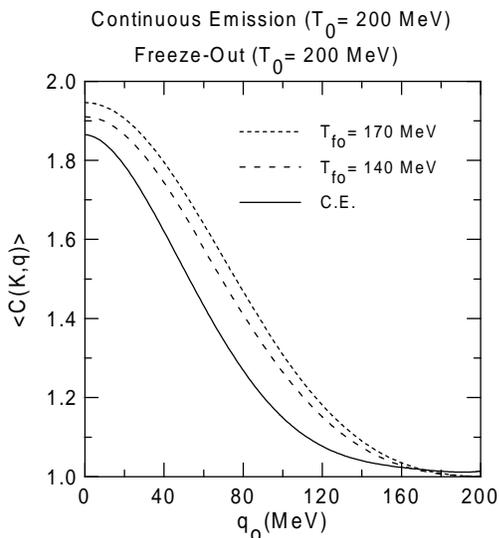}} 
\caption{\it \small $\langle $ C $\rangle _{\pi \pi }$ {\sl x} $%
q_O$ averaged over $q_L$ and $q_S$, showing the curve corresponding to the
continuous emission hypothesis as compared to the ones corresponding to
usual freezeout at $T_{fo}=170$ MeV and $T_{fo}=140$ MeV, but identical
initial temperatures $T_0$.}
\label{fig5.eps}
\end{figure}

\vspace{.3cm}
\begin{figure}[h]\epsfxsize=8cm
\centerline{\epsfbox{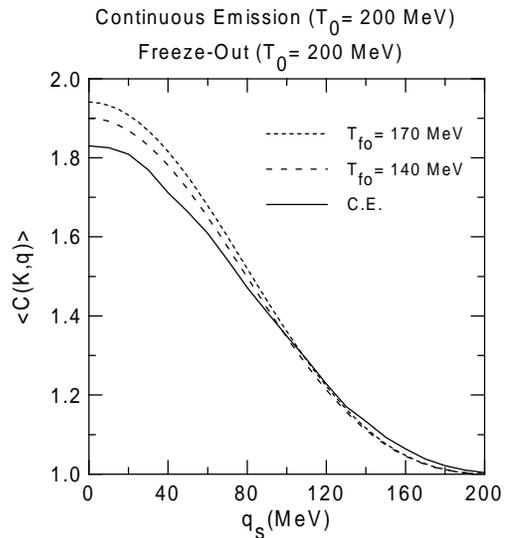}} 
\caption{\it \small $\langle$ C $\rangle_{\pi \pi}$ {\sl x} $q_S$ 
averaged over $q_L$ and $q_O$, showing the curve corresponding to the 
continuous emission hypothesis as compared to the ones corresponding to 
usual freezeout at $T_{fo}=170$ MeV and $T_{fo}=140$ MeV, but identical 
initial temperatures $T_0$.} 
\label{fig6.eps}
\end{figure}

\noindent
$q_S\,$, the
depletion of $\langle$ C $\rangle_{\pi \pi}$ is more pronounced in CEM as
compared to the curves corresponding to the sharp freezeout case. As
happened in the previous cases, for the {\bf same} initial temperature, 
$\langle C\rangle _{\pi \pi }$ with lower $T_{fo}\,$ is closer to the 
curve for CEM, but the shape is somewhat different.

In the previous figures, we have shown and discussed the differences of 
CEM correlation function, confronted with the usual 
abrupt freezeout one, when the fluid started from the same initial 
conditions. Now, when analyzing the experimental data, the parameters are 
usually adjusted by fitting the data points as close as possible, and the 
conclusions are extracted from the adjusted parameters. In the present 
model calculations, the only parameter, besides the freezeout temperature 
$T_{fo}\,$, is the initial temperature $T_0$. For computing the results 
presented in Figs. 7-9, we have fixed the initial temperature for CEM as 
200 MeV, and varied the initial temperature $T_0$ for the usual freezeout 
scenario, trying to get the same (or similar) result. To doing so, we have 
chosen the freezeout temperature as $T_{fo}=140$ MeV, as often done in 
hydrodynamic calculations and following the indications of the previous 
discussions. 

As seen, especially in Fig. 7, the correlation functions predicted by the
two different scenarios were so different in shape that it was not always
possible to obtain similar curves. For example, in the case of $q_L$
dependence, Fig. 7, $\langle$C$\rangle$ for CEM is closer to the curve 
with higher $T_0$ at small $q_L$, but at high $q_L$, it turns to be 
closer to the one with the lowest value of $T_0$. 
As discussed above in connection with Fig. 4, the correlation curve in 
CEM could be interpreted as showing the history of the hot matter in 
expansion, {\it i.e.}, the tail of $\langle$C$\rangle$ reflects 
essen-\hfilneg\ 

\newpage
\vspace{.4cm}

\begin{figure}[h]\epsfxsize=8cm 
\centerline{\epsfbox{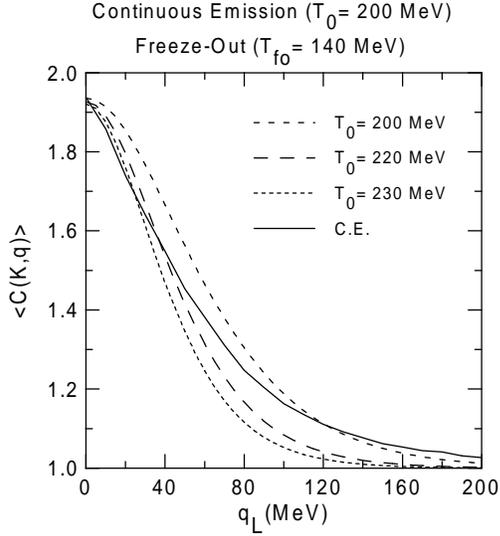}} 
\caption{\it \small $\langle$ C $\rangle_{\pi \pi}$ {\sl x} $q_L$ averaged 
over $q_S$ and $q_O$, showing the curve corresponding to the continuous 
emission hypothesis with the ones corresponding to the usual freeze-out at 
$T_{fo}=140$ MeV, but different initial temperatures $T_0$.} 
\label{fig7.eps}
\end{figure}

\noindent 
tially 
the early times, when the size of the fluid is small and its temperature 
high, and the peak the later times, when the fluid has fully expanded and 
cooled down. 
In the usual freezeout picture with a fixed $T_{fo}\,$, small $T_0$ is 
enough to produce a large tail, whereas a larger expansion, so higher  
$T_0\,$, is required to produce a narrower $\langle$C$\rangle$.

In Fig. 8, where the $q_O$ dependence of $\langle C\rangle_{\pi \pi}$ is 
shown, we can again observe a distortion introduced by CEM into the shape 
of the correlation function. We see that $\langle C\rangle$ for CEM is 
closer to the curve with $T_0=230$ MeV\hfilneg\ 

\begin{figure}[h]\epsfxsize=8cm
\centerline{\epsfbox{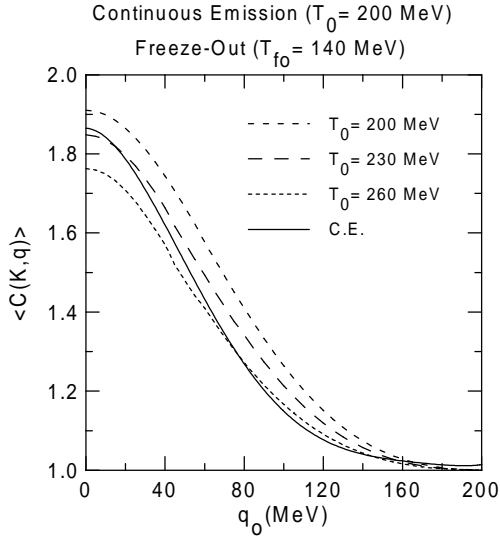}} 
\caption{\it \small $\langle$ C $\rangle_{\pi \pi}$ {\sl x} $q_O\,$, 
averaged over $q_L$ and $q_S$, showing the curve corresponding to CEM as 
compared to the ones corresponding to the sharp freezeout at $T_{fo}=140$ 
MeV, but different initial temperatures $T_0$.} 
\label{fig8.eps}
\end{figure}

\noindent 
at small $q_O$, but at high $q_O$, 
it turns to be closer to the one corresponding to $T_0=260$ MeV. However, 
differently from the previous case, Fig. 7, the shape of the freezeout 
correlation curves is only slightly dependent on $T_0$ and it becomes 
narrower as the temperature increases. What is clearly $T_0$ dependent 
here is the intercept at the origin, which reflects the large expansion 
dependence of $\langle C\rangle_{\pi \pi}$ {\sl x} $q_L$ as shown in 
Figs. 4 and 7.

Finally, Fig. 9 shows that in this case it is possible to find an
appropriate $T_0$ for sharp freezeout to reproduce the CEM curve. We see
that $\langle$ C $\rangle$ for CEM is closer to the curve with 
$T_0=230$ MeV and the agreement is good for most of the $q_S$ region 
where the interferometric signal is present. This was expected because in 
the present study we neglected the transverse expansion, so the 
transverse size is the same in both the scenarios. The initial 
temperature $T_0$ in freezeout is higher than the one for CEM, because 
this is required to make the size of the fluid large enough and the 
correlation in the longitudinal direction sharp enough to decrease the 
intercept on averaging. 


\begin{figure}[h]\epsfxsize=8cm
\centerline{\epsfbox{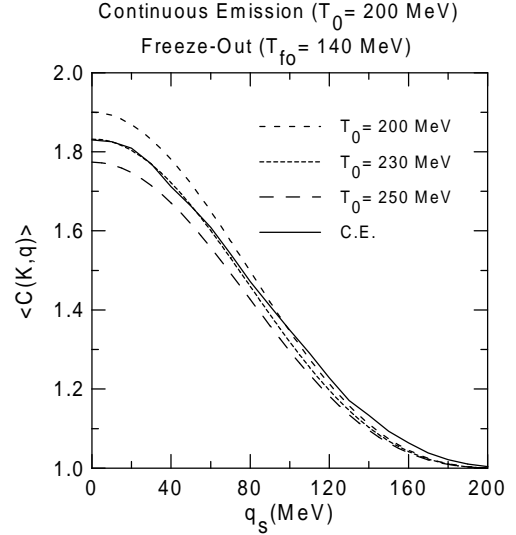}} 
\caption{\it \small $\langle$ C $\rangle_{\pi \pi}$ {\sl x} $q_S$ 
averaged over $q_L$ and $q_O$, showing the curve corresponding to CEM as 
compared to the ones corresponding to usual freezeout at $T_{fo}=140$ MeV, 
but with different initial temperature $T_0\,$.} 
\label{fig9.eps}
\end{figure}

From the above results, mainly from the correlation curves as function of 
$q_L\,$, we clearly see deviations from the pure Gaussian behavior in 
cases where the continuous emission ansatz was assumed. This could 
actually reflect a signature of a continuous process in the particle 
emission. 

\section{Discussions and Concluding Remarks}

As mentioned in the Introduction, treating the decoupling process in
heavy-ion collisions as occurring on a sharply defined surface is an
operationally simple but highly idealized description. If the consideration
of a finite thickness of such a decoupling region does not bring any
noticeable difference in the observable quantities, such an approximation
would be unquestionable. However, previous studies\cite{ghk,ghks,gs1,gs2}
have shown that several quantities, such as transverse spectra of produced
particles and heavy-particle production ratios are sensitive to more
involved description of the process, called continuous emission model\cite
{ghk}.

In this paper, we concentrated on the two-pion interferometry, which has 
extensively been used as a powerful tool for extracting the space-time 
geometry, as well as probing the underlying dynamics of the hadronic 
matter formed in heavy-ion collisions, and studied the differences 
introduced by CEM in confront with the usual sudden freezeout. As shown 
in Sec. IV, also the HBT effect suffers a large deformation when the 
usual freezeout is replaced by CEM. This means that conclusions 
achieved on the properties of the matter formed in high-energy 
collisions may differ substantially if we adopt one or the other 
scenario studied here. 

For the sake of conceptual clarity and, evidently, also to simplify the
computation, we have adopted in this work a simplified one-dimensional 
Bjorken model for massless pion fluid (the pion mass has been included 
only to computing observable quantities), without phase transition. 
Nevertheless, one general result emerges, which seems to be evident 
especially by looking at Figs. 7-9. Namely, if we describe the same data 
by using CEM or sharp freeze-out (with $T_{fo}=140$ MeV), the initial 
temperature $T_0\,$ required in CEM is lower than in the usual 
freeze-out. 
If $T_{fo}$ is higher, the difference in $T_0$ becomes even larger, which 
is clear from Figs. 4-6. This result means that if CEM is the correct 
description of the decoupling process, then it is harder to reach the 
quark-gluon plasma phase than it appears in the usually adopted sharp 
freezeout scenario.

Since we have worked with a simplified model, we did not attempt to make 
any comparison with data. For doing this, evidently we have to do some 
(or all) of the following improvements. More realistic equation of state 
(probably including phase transition) should be used; finite longitudinal 
extension of the fluid should be considered; transverse expansion should 
be included; resonance formation should be taken into account. All these 
modifications require some hydrodynamic numerical code with CEM 
incorporated. We are now working in this direction.

\bigskip
\bigskip

\noindent{\bf Acknowledgement}

This work was partially supported by 
FAPESP 
(contract nos. 1998/02249-4 and 1998/14990-0) and by 
CNPq 
(contract no. 300054/92-0). We express our gratitude to T. Kodama and 
T. Cs\"org\H o for stimulating discussion on the results.

\end{document}